\documentclass[a4paper,11pt]{article}

\usepackage[utf8]{inputenc}
\usepackage{amssymb,amsfonts,graphicx,color,subfig,ae}
\usepackage[english]{babel}
\usepackage[intlimits]{amsmath}
\usepackage{eucal}
\usepackage{cite}
\usepackage{hyperref}
\usepackage[heightrounded]{geometry}

\newcommand{\2}{{\textstyle \frac{1}{2}}}
\newcommand{\mcH}{\mathcal{H}}

\newcommand{\mcS}{\mathcal{S}}

\DeclareMathOperator{\tr}{tr}
\DeclareMathOperator{\trf}{\bf trf}
\DeclareMathOperator{\sh}{sh}
\DeclareMathOperator{\ch}{ch}
\DeclareMathOperator{\tgh}{th}
\DeclareMathOperator{\cth}{cth}
\DeclareMathOperator{\sign}{sign}
\DeclareMathOperator{\dn}{dn}

\newcommand{\dx}[1]{\mathrm{d}#1}

\newcommand{\mfa}{\mathfrak{a}}

\newcommand{\mfb}{\mathfrak{b}}

\newcommand{\mfbq}{\bar{\mathfrak{b}}}

\newcommand{\opv}{\mathbf{v}}

\newcommand{\mut}{\tilde{\mu}}
\newcommand{\s}{\sigma}

\newcommand{\epp}{\, .}
\newcommand{\epc}{\, ,}

\title{Short-distance thermal correlations in the massive XXZ chain}

\author{
Christian Trippe\thanks{e-mail:
\href{mailto:trippe@physik.uni-wuppertal.de}%
{\protect\nolinkurl{trippe@physik.uni-wuppertal.de}}}\;,
Frank G\"ohmann\thanks{e-mail:
\href{mailto:goehmann@physik.uni-wuppertal.de}%
{\protect\nolinkurl{goehmann@physik.uni-wuppertal.de}}}\;,
Andreas Kl\"umper\thanks{e-mail:
\href{mailto:kluemper@physik.uni-wuppertal.de}%
{\protect\nolinkurl{kluemper@physik.uni-wuppertal.de}}}
\\
\parbox{0.9 \linewidth}{\vspace{0.4 \baselineskip}\centering
    Fachbereich C -- Physik, Bergische Universit\"at Wuppertal,\\
    42097 Wuppertal, Germany}
}

\date{\today}

\begin{document}

\maketitle

\begin{abstract}
We explore short-distance static correlation functions in the infinite
XXZ chain using previously derived formulae which represent the
correlation functions in factorized form. We compute two-point functions
ranging over 2, 3 and 4 lattice sites as functions of the temperature
and the magnetic field in the massive regime $\Delta>1$, extending our
previous results to the full parameter plane of the antiferromagnetic
chain ($\Delta > -1$ and arbitrary field $h$). The factorized formulae
are numerically efficient and allow for taking the isotropic limit
($\Delta = 1$) and the Ising limit ($\Delta = \infty$). At the critical
field separating the fully polarized phase from the N\'eel phase,
the Ising chain possesses exponentially many ground states. The
residual entropy is lifted by quantum fluctuations for large but finite
$\Delta$ inducing unexpected crossover phenomena in the correlations.
\end{abstract}

\section{Introduction}
All observable information about a many-body quantum system
is encoded in its correlation functions. For interacting systems in
the thermodynamic limit this information is, in general, only accessible
through various approximations.\footnote{For a recent perturbative approach 
to the dynamical correlation functions of the XXZ chain in the massive
regime see \cite{JGE09}.} Until rather recently very few examples were
known, where correlation functions could be calculated exactly. Among
the known examples there was none with short-range interactions of
finite and tunable strength. In particular, for the large and prototypical
class of integrable models that are solvable by Bethe ansatz, the
solution was limited to the spectral properties, but except for what can
be concluded from the finite size corrections to the spectrum rather
little was known about their correlation functions.

This situation is about to change. Extensive studies by several groups,
mainly focusing on the XXZ quantum spin chain, have eventually uncovered a
structure which we believe to be typical for the Bethe ansatz solvable
models: the one-point and the neighbour correlation functions
determine everything. We call this phenomenon factorization. Based on
the factorization we can calculate for the first time short-range
correlation functions with arbitrary accuracy, both, for finite
temperatures in the thermodynamic limit \cite{BDGKSW08}, and for arbitrary
finite length in the ground state \cite{DGHK07}. This has applications
e.g.\ in the determination of ESR-line shifts \cite{MSO05} and is useful
for estimating the errors associated with standard numerical techniques
like the quantum Monte Carlo algorithm \cite{Loew07}.

Factorization appeared first \cite{BoKo01} as the explicit factorization
of certain multiple integrals \cite{JMMN92,JiMi96,KMT99b} describing
the density matrix elements of the isotropic Heisenberg chain in the
ground state with no magnetic field applied. A similar factorization
for the ground state correlation functions of the XXZ chain in the 
massive regime, i.e.\ for the model considered in this article,
was carried out in \cite{TKS04,KSTS04}. After a multiple integral
representation for finite temperature correlation functions
had become available \cite{GKS04a,GKS05}, the factorization of the
integrals for short distances was extended to this case \cite{BGKS06}.
Following the success with the direct factorization of the integrals a deep
exploration into the algebraic structure of the static correlation
functions of the XXZ chain was conducted by Boos, Jimbo, Miwa, Smirnov and
Takeyama, culminating in the works \cite{BJMST08a,BJMS09app,JMS08,%
BoGo09}, where the factorization of all static correlation functions
of the XXZ chain was proved in a very general setting, including the
finite temperature and finite magnetic field cases.

In all the recent works cited above it was crucial not to deal directly
with the spin chain Hamiltonian, but with the associated six-vertex
model \cite{BaBook}. The six vertex model naturally carries inhomogeneity
parameters in horizontal and vertical directions and can be distorted by a
disorder field without loosing its distinctive feature, the integrability.
In this setting the density matrix of a finite segment of the
XXZ chain is naturally generalized to include the inhomogeneity
parameters and the strength $\alpha$ of the disorder field. The vertical
inhomogeneity parameters $\nu_j$ and the disorder field regularize the
expression for the density matrix. The horizontal spectral parameters
allow one to introduce the temperature into the model \cite{GKS04a,%
Kluemper92, Kluemper93} and to adjust the Hamiltonian \cite{TrKl07}. The
density matrix of the inhomogeneous six-vertex model with disorder field
depends polynomially on only two complex functions $\varphi (\nu|\alpha)$
and $\omega(\nu_1, \nu_2|\alpha)$ which are efficiently described in terms
of the solutions of integral equations \cite{BoGo09}. The calculation of
the coefficients of the polynomials is a purely algebraic problem. It is
related to the construction of a special fermionic basis \cite{BJMST08a,%
BJMS09app} on the space of local operators acting on the space of states of
the spin chain. We call this the algebraic part of the problem and
the calculation of the functions $\varphi$ and $\omega$ the physical
part of the problem, since all dependence on the physical parameters,
like temperature magnetic field or boundary conditions, is in these
two functions.

For the time being no efficient algorithm for the algebraic part of the
problem is known. This limits the range of the correlation functions we
are actually able to calculate to a few lattice sites. In \cite{BDGKSW08}
we calculated the coefficients for the two-point functions for up to four
lattice site by brute force computer algebra. Then we solved the physical
part of the problem in the massless regime. At a late stage of the
calculation the homogeneous limit and the limit of vanishing disorder
field had to be taken in order to obtain the correlation functions
of the spin chain. Due to the singular nature of this limit the first
derivative of $\omega$ with respect to $\alpha$ and various derivatives
with respect to the inhomogeneity parameters appear. The final expressions
are polynomials in three functions $\omega$, $\omega'$ (basically the
$\alpha$ derivative of $\omega$) and $\varphi$ and their derivatives.
The coefficients in these polynomials are rational functions in the
deformation parameter $q$, which determines the anisotropy of the
XXZ chain, and are algebraic and universal. Therefore the expressions
derived in \cite{BDGKSW08} for the two-point functions in the
massless regime also apply in the massive regime, if one replaces
the physical part, $\omega$, $\omega'$ and $\varphi$ appropriately.

This is the main subject of this work. We reformulate the expressions
for the functions $\omega$, $\omega'$ and $\varphi$ originally obtained
in \cite{BGKS07} in a way that is convenient for numerical calculations
in the massive regime $\Delta > 1$ of the XXZ chain. Combining this with the
general formulae and the numerical results from \cite{BDGKSW08} we obtain
accurate data for the two-point functions in the full parameter plane of
the infinite antiferromagnetic chain (anisotropy $\Delta > - 1$ and
magnetic field $h$ arbitrary). They exhibit a surprisingly rich
non-monotonic behaviour.

We should comment on the level of mathematical rigour of our results.
In \cite{BGKS07,BDGKSW08} we conjectured the physical part of the
problem as well as the contribution to the algebraic part connected
with the one-point functions in the limit, when $\alpha \rightarrow 0$.
Meanwhile we know \cite{JMS08,BoGo09} the exact physical
part even for finite $\alpha$ and we could show \cite{BoGo09} that
it reproduces our conjecture for $\alpha \rightarrow 0$. The algebraic
part for distances up to four lattice sites was checked in
\cite{BDGKSW08} by several independent means (comparison with the
high temperature expansion of the multiple integrals \cite{GKS05},
direct numerical computation, consideration of various limits). Now,
as we may infer from \cite{BJMST08a,JMS08,BoGo09}, it is clear
that the exponential formula of \cite{BDGKSW08} is valid for arbitrary
distances and zero magnetic field and at least for the distance
of two lattice sites at finite magnetic field. On the other hand,
the work \cite{JMS08} offers a future way for the rigorous calculation
of the algebraic part which needs no exponential form of the density
matrix and will hopefully lead to the calculation of static correlation
functions for points at larger distances.

\section{Hamiltonian and density matrix}
We consider the Hamiltonian
\begin{equation} \label{ham}
     \mcH_N=J\sum_{j=-N+1}^N
        \left(\sigma_{j-1}^x\sigma_{j}^x+\sigma_{j-1}^y\sigma_{j}^y
        +\Delta(\sigma_{j-1}^z\sigma_{j}^z-1)\right)
\end{equation}
of the XXZ chain in the massive antiferromagnetic regime ($J>0$ and
$\Delta>1$). The $\sigma_j^\alpha, \, j=-N+1,\ldots, N$, act locally as
Pauli matrices, $\Delta=\ch(\eta)$ is the anisotropy para\-meter and $J$
the exchange coupling.

The XXZ Hamiltonian preserves the $z$-component of the total spin
\begin{equation}
     \mcS_N^z=\frac{1}{2}\sum_{j=-N+1}^N \sigma_{j}^z .
\end{equation}
Thus, the magnetization in $z$-direction is a thermodynamic quantity,
and the thermal equilibrium of the finite system is characterized by the
statistical operator
\begin{equation} \label{defrhon}
     \rho_N(T,h)=\frac{e^{-(\mcH_N-h\mcS_N)/T}}
                      {\tr_{-N+1,\ldots,N}e^{-(\mcH_N-h\mcS_N)/T}}
\end{equation}
depending on the temperature $T$ and the external magnetic field $h$.

In the thermodynamic limit $L=2N\rightarrow\infty$ the system severely
simplifies, since, from the six-vertex model point of view, a single
eigenstate of the so-called quantum transfer matrix determines the state
of thermodynamic equilibrium and hence all static correlation functions
\cite{GKS04a}. Clearly the naive limit $N \rightarrow \infty$ makes
no sense in (\ref{defrhon}). To perform this limit in a sensible way we
fix integers $m, n$ with $m < n$ and introduce the density matrix
\begin{equation}
     D_{[m,n]} (T,h)=\lim_{N\rightarrow\infty}
              \tr_{-N+1,\ldots,m-1,n+1,n+2,\ldots,N}\rho_N(T,h)
\end{equation}
of the segment $[m,n]$ of the infinite chain which is well defined.
It has the reduction properties
\begin{equation} \label{densredu}
     \tr_m D_{[m,n]} (T,h) = D_{[m+1,n]} (T,h) \epc \quad
        \tr_n D_{[m,n]} (T,h) = D_{[m,n-1]} (T,h) \epp
\end{equation}

We consider the vector space $\cal W$ of operators on the infinite
chain which act non-trivially only on a finite number of lattice sites
($\sigma_j^z \sigma_k^z$ for arbitrary fixed $j$ and $k$ is an example
of such an operator). On this space we define a map $\trf_{[m,n]}:
{\cal W} \rightarrow \bigl( {\mathbb C}^2 \bigr)^{\otimes (n - m + 1)}$ by
\begin{equation}
     \trf_{[m,n]} {\cal O} = \dots \2 \tr_{m - 2} \2 \tr_{m - 1}
                                   \2 \tr_{n + 1} \2 \tr_{n + 2} \dots
                             {\cal O} \epp
\end{equation}
It restricts the action of ${\cal O}$ to the interval $[m,n]$.

Because of (\ref{densredu}) the definition $Z: {\cal W} \rightarrow
{\mathbb C}$,
\begin{equation}
     Z({\cal O}) =
        \lim_{\substack{m \rightarrow - \infty \\ n \rightarrow \infty}}
        \tr_{m, \dots, n} D_{[m,n]} (T,h) \trf_{[m,n]} {\cal O}
\end{equation}
makes sense and determines the thermal expectation value of ${\cal O}$.
The functional $Z$ has the natural interpretation of the statistical
operator on the space $\cal W$ and may be thought of as the thermodynamic
limit of $\rho_N$, equation (\ref{defrhon}).

The functional $Z$ allows for a generalization within the framework
of the six-vertex model with disorder field. In the seminal paper
\cite{JMS08} this generalization was denoted $Z^\kappa$ (with $\kappa$
referring to the magnetic field in certain units). It is this functional
which depends on only two functions $\varphi (\nu|\alpha)$ and
$\omega(\nu_1, \nu_2|\alpha)$ and which clearly separates the physical
and the algebraic part of the problem. For details we refer the reader
to \cite{JMS08,BoGo09}.

In this work we are dealing with applications. We shall only need an
understanding of the general idea of factorization and the concrete
formula obtained in \cite{BDGKSW08}. For this reason we refrain in the
following from any sophisticated mathematical notation and simply write
$\langle {\cal O} \rangle_{h, T} = Z({\cal O})$ for thermal expectation
values on the infinite chain.
%\comment{Something more about the construction of the function $\omega$
%and the correlation functions.}
%
%The algebraic part of the construction for the massive regime is the
%same as for the critical, which is described in \cite{BDGKSW08}.

\section{The physical part of the construction}
\label{sec:physpart}
As explained in \cite{BDGKSW08} the limit $\alpha \rightarrow 0$ leads
to three functions $\omega(\mu_1, \mu_2)$, $\omega'(\mu_1,\mu_2)$ and
$\varphi(\mu)$ that determine the correlation functions of the XXZ chain.
A relatively simple description of these functions in terms of an
auxiliary function $\mfa$ and a generalized magnetization density $G$ was
given in \cite{BGKS07}. We call it the $\mfa$-formulation. The
$\mfa$-formulation is useful for deriving multiple integral formulae
\cite{GKS04a} and for studying the high-temperature expansion of the
free energy and the correlation functions \cite{TsSh05}. It allows
for a rather uniform description of the massless and the massive XXZ
chain, the only difference between the two cases being a different choice
for the so-called canonical integration contour in the complex plane
\cite{GKS04a}.

For numerical calculations of thermodynamic properties
\cite{Kluemper92,Kluemper93} or short-distance correlators from the
multiple integral \cite{BoGo05,BDGKSW08} one has to change to a different
formulation we refer to as the $\mfb\mfbq$-formulation. It needs pairs
of functions, but the defining integral equations involve only straight
contours. It is here where one has to distinguish between the massless and
the massive case, when it comes to numerical calculations. Two
separate computer programs are needed. In \cite{BDGKSW08} we studied
the massless case. Here we proceed to the massive case. We start
from the $\mfa$-formulation of the physical part proposed in \cite{BGKS07}
and switch to the $\mfb\mfbq$-formulation. This is a standard
procedure which (in the massive case) basically requires the application
of Fourier series and the convolution theorem. As far as the integral
equations are concerned the reader may e.g.\ refer to \cite{BoGo05}.
Then it is rather obvious how to proceed in a similar way with the
expressions for the functions $\varphi$, $\omega$ and $\omega'$
(equations (15), (17) and (20) for $\alpha = 0$ in \cite{BGKS07}). For
this reason we just state the results.

Let us define a pair of auxiliary functions $\mfb$, $\mfbq$ as the solution
of the non-linear integral equations (NLIE)
% \begin{align}
% \ln\mfb(x)=&-\frac{h}{2T}-\frac{2J\sh(\eta)}{T}d(x)
% +\int_{-\pi/2}^{\pi/2}\frac{\dx{y}}{\pi}\kappa(x-y)\ln\left(1+\mfb(y)
% \right)-\int_{-\pi/2}^{\pi/2}\frac{\dx{y}}{\pi}\kappa(x-y-i\eta^+)\ln
% \left(1+\mfbq(y)\right)\\
% \ln\mfbq(x)=&\frac{h}{2T}-\frac{2J\sh(\eta)}{T}d(x)
% +\int_{-\pi/2}^{\pi/2}\frac{\dx{y}}{\pi}\kappa(x-y)\ln\left(
% 1+\mfbq(y)\right)-\int_{-\pi/2}^{\pi/2}\frac{\dx{y}}{\pi}
% \kappa(x-y+i\eta^-)\ln\left(1+\mfb(y)\right)
% \end{align}
\begin{subequations}
\label{eq:nlie}
\begin{align}
     \ln\mfb(x) = &-\frac{h}{2T} - \frac{2J\sh(\eta)}{T}d(x) 
                  + \kappa\ast\ln\left(1+\mfb\right) (x)
                  - \kappa^-\ast\ln\left(1+\mfbq\right) (x) \epc \\[1ex]
     \ln\mfbq(x) = &\frac{h}{2T} - \frac{2J\sh(\eta)}{T}d(x)
                  + \kappa\ast\ln\left(1+\mfbq\right) (x)
                  - \kappa^+\ast\ln\left(1+\mfb\right) (x)
\end{align}
\end{subequations}
with $f\ast g(x)=\frac{1}{\pi}\int_{-\pi/2}^{\pi/2}\dx{y}f(x-y)g(y)$
denoting a convolution.

The integral equations are specified by the integration kernels $\kappa$,
$\kappa^\pm$ and the driving terms, which contain a single transcendental
function $d$. Note that the physical para\-meters temperature $T$, coupling
$J$ and magnetic field $h$ enter into the calculation of the correlation
functions only through the driving terms of the NLIE \eqref{eq:nlie}.

The functions $d$ and $\kappa$, $\kappa^\pm$, as well as some other
functions occurring in other integral equations below, have simple
Fourier series representations which arise naturally in the derivation
of the non-linear equations and which are also useful for solving them
numerically,
\begin{subequations}
\begin{align}
     d(x) & = \sum_{k=-\infty}^\infty \frac{e^{i2kx}}{\ch(\eta k)}
              \displaybreak[0] \epc \\
     \kappa(x) & = \sum_{k=-\infty}^\infty
                   \frac{e^{-\eta|k| + 2ikx}}{2\ch(\eta k)}
                   \epc \quad
     \kappa^\pm(x) = \kappa(x\pm i\eta^\mp) \epp
\end{align}
\end{subequations}

Alternatively the kernel and the function $d$ can be realized in terms
of special functions. For the function $d$ we find 
\begin{equation}
\label{eq:d-funct}
     d(x)=\frac{2K}{\pi}\dn\left(\frac{2Kx}{\pi},i\frac{\eta}{\pi}\right)
          \epc
\end{equation}
where $\dn(x,\tau)$ is one of the Jacobi elliptic functions \cite{AbSt72},
and $K$ is the complete elliptic integral of the first kind.

The integration kernel $\kappa$ can be expressed in terms of the normalized
trigonometric gamma function $T$ introduced in \cite{Ruijsenaars97}.
In order to define it we first of all recall the definition of
the $q$-gamma function \cite{GaRa90},
\begin{equation}
     \Gamma_q (z) = (1-q)^{1-z} \prod_{n=1}^\infty \frac{1-q^n}{1-q^{z+n+1}}
                    \epp
\end{equation}
Then, for $q=\exp(-4\eta)$,
\begin{equation}
     T(r;z) = \frac{\Gamma_q(1/2)}{\Gamma_q(iz+1/2)}
              \exp\left(-\frac{1}{2}\ln\pi
              + \frac{rz^2}{2} - iz\ln\left(\frac{1-e^{-2r}}{2r}\right)
                \right)
\end{equation}
and
\begin{equation}
     \kappa(x) = \frac{1}{2i} \partial_x
                 \ln \left[ \frac{T\bigl(2\eta;\frac{x}{2\eta}\bigr)
                                  T\bigl(2\eta;-\frac{x}{2\eta}-\frac{i}{2}
                                  \bigr)}
                                 {T\bigl(2\eta;-\frac{x}{2\eta}\bigr)
                                  T\bigl(2\eta;\frac{x}{2\eta}
                                  -\frac{i}{2}\bigr)}
                     \right] \epp
\end{equation}

In addition to the auxiliary functions $\mfb$ and $\mfbq$ we need two
more pairs of functions $g^\pm_\mu$ and ${g'}^\pm_\mu$ to express
$\omega$, $\omega'$ and $\varphi$ by means of auxiliary functions. Both
are solutions of linear integral equations involving $\mfb$ and $\mfbq$,
\begin{subequations}
\begin{align}
     g^+_\mu(x) & = -d(x-\mu) + \kappa\ast\frac{g^+_\mu}{1+\mfb^{-1}}(x)
                    - \kappa^-\ast\frac{g^-_\mu}{1+\mfbq^{-1}}(x) \epc
                    \\[1ex]
     g^-_\mu(x) & = -d(x-\mu) + \kappa\ast\frac{g^-_\mu}{1+\mfbq^{-1}}(x)
                    - \kappa^+\ast\frac{g^+_\mu}{1+\mfb^{-1}}(x)
\end{align}
\end{subequations}
and
\begin{subequations}
\label{eq:gprime}
\begin{align} \notag
     {g'}^+_\mu(x) = & -\eta c_+(x-\mu) 
                       + \eta l\ast\frac{g^+_\mu}{1+\mfb^{-1}}(x)
                       - \eta l^-\ast\frac{g^-_\mu}{1+\mfbq^{-1}}(x)\\
                     & \mspace{144.mu}
                       + \kappa\ast\frac{{g'}^+_\mu}{1+\mfb^{-1}}(x)
                       - \kappa^-\ast\frac{{g'}^-_\mu}{1+\mfbq^{-1}}(x)
                     \epc \\[2ex] \notag 
     {g'}^-_\mu(x) = & - \eta c_{-}(x-\mu)
                       + \eta l\ast\frac{g^-_\mu}{1+\mfbq^{-1}}(x)
                       - \eta l^+\ast\frac{g^+_\mu}{1+\mfb^{-1}}(x)\\
                     & \mspace{144.mu}
                       + \kappa\ast\frac{{g'}^-_\mu}{1+\mfbq^{-1}}(x)
                       - \kappa^+\ast\frac{{g'}^+_\mu}{1+\mfb^{-1}}(x) \epp
\end{align}
\end{subequations}
The new integration kernel $l$ and the new functions $c^\pm$ occurring
in (\ref{eq:gprime}) are conveniently described as Fourier series,
\begin{equation}
     l(x) = \sum_{k=-\infty}^\infty \frac{\sign(k) e^{i2kx}}{4\ch^2(\eta k)}
            \epc \quad l^\pm(x) = l(x\pm i\eta^\mp) \epc
\end{equation}
where we used the standard convention $\sign(0)=0$, and
\begin{equation}
     c_\pm(x) = \pm\sum_{k=-\infty}^\infty
                \frac{e^{\pm\eta k + 2ikx}}{2\ch^2(\eta k)} \epp
\end{equation}

The functions $\omega(\mu_1,\mu_2)$, $\omega'(\mu_1,\mu_2)$ and
$\varphi(\mu)$ which determine the inhomogeneous correlation functions
can be expressed as integrals over the above functions.
The function
\begin{equation} \label{eq:phi}
     \varphi(\mu) = \int_{-\pi/2}^{\pi/2} \frac{\dx{x}}{2\pi}
                     \left(
                     \frac{g_{\mut}^-(x)}{1+\mfbq(x)^{-1}}
		      -\frac{g_{\mut}^+(x)}{1+\mfb(x)^{-1}}
		      \right)
\end{equation}
with\footnote{In the following this convention will also be used
for $\mut_1,\mut_2$.} $\mut=-i\mu$ is related to the magnetization
\begin{equation}
     m(h,T) = \frac{1}{2}\left\langle\sigma_j^z\right\rangle_{T, h}
            = - \frac{1}{2} \varphi(0)
\end{equation}
which is the only independent one-point function of the XXZ chain.
The function
\begin{equation}
     \omega(\mu_1,\mu_2) = - 4\kappa(\mut_2-\mut_1)
        + \tilde{K}_\eta(\mut_2-\mut_1)
        - d\ast\left(\frac{g_{\mut_1}^+}{1+\mfb^{-1}}
        + \frac{g_{\mut_1}^-}{1+\mfbq^{-1}}\right)(\mut_2)
\end{equation}
with
\begin{equation}
     \tilde{K}_\eta(x) = \frac{\sh(2\eta)}{2\sin(x+i\eta)\sin(x-i\eta)}
\end{equation}
also determines the internal energy \cite{BDGKSW08}. The last function
which is necessary to determine the correlation functions is
\begin{multline}
     \omega'(\mu_1,\mu_2) = -4\eta l(\mut_2-\mut_1)
        - \eta \tilde{L}_\eta(\mut_2-\mut_1)
        - d\ast\left(\frac{{g'}_{\mut_1}^+}{1+\mfb^{-1}}
        + \frac{{g'}_{\mut_1}^-}{1+\mfbq^{-1}}\right)(\mut_2)\\
        - \eta c_{-}\ast\frac{g_{\mut_1}^+}{1+\mfb^{-1}}(\mut_2)
        - \eta c_{+}\ast\frac{g_{\mut_1}^-}{1+\mfbq^{-1}}(\mut_2) \epc
\end{multline}
where
\begin{equation}
     \tilde{L}_\eta(x) = \frac{i\sin(2x)}{2\sin(x+i\eta)\sin(x-i\eta)} \epp
\end{equation}
The physical meaning of $\omega'$ is less intuitive. Its derivative
appears in the two-point functions and is therefore related to certain
neighbour correlators.

As shown in \cite{BGKS07,BDGKSW08} it is necessary to perform the
homogeneous limit for the calculation of correlation functions. This
limit seems rather singular, because the coefficients coming from the
algebraic part in general have poles, when two inhomogeneities coincide.
Still, these poles are canceled by zeros coming from certain symmetric
combinations of the functions $\omega (\mu_1, \mu_2)$,
$\omega' (\mu_1, \mu_2)$ and $\varphi (\mu)$ in the numerators. In order
to perform the limit one has to apply l'H\^ospital's rule. This finally
leads to polynomials in the functions $\omega(0,0)$, $\omega'(0,0)$,
$\varphi(0)$ and in derivatives of these functions with respect to the
inhomogeneity parameters evaluated at zero. We denote derivatives with
respect to the first argument by subscripts $x$ and derivatives with
respect to the second argument by subscripts $y$ and leave out zero
arguments for simplicity. Then e.g.\ $\omega_{xyy} = \partial_x
\partial_y^2 \omega (x,y)|_{x,y = 0}$ etc.

For the examples in the next section the non-linear integral equations
for $\mfb$ and $\mfbq$ as well as their linear counterparts for
$g_\mu^{(\pm)}$ and ${g'}_\mu^{(\pm)}$ were solved iteratively in Fourier
space, using the fast Fourier transformation algorithm. The derivatives
of $g_\mu^{(\pm)}$ and ${g'}_\mu^{(\pm)}$ with respect to $\mu$, needed
in the computation of the respective derivatives of $\omega$ and $\omega'$, 
satisfy linear integral equations as well, which were
obtained as derivatives of the equations for $g_\mu^{(\pm)}$ and
${g'}_\mu^{(\pm)}$. The modifications due to the derivatives are
particularly simple in Fourier space.

\section{Examples of short distance correlators}
\label{sec:examples}
In the sequel we shall restrict ourselves to the longitudinal and
transversal two-point functions $\left\langle \sigma_1^z
\sigma_n^z\right\rangle _{T,h}$, $\left\langle \sigma_1^x
\sigma_n^x\right\rangle _{T, h}$ for $n = 2, 3, 4$.\footnote{Our method is based on the density matrix and allows us to obtain all its matrix elements. In particular, we may calculate all other independent two-point functions like e.g. $\left\langle \sigma_1^x \sigma_n^y\right\rangle _{T,h}$ for $n= 2, 3, 4$. Still, since this will give more bulky expressions of the type \eqref{szsz4}, \eqref{sxsx4}, we refrained from the temptation of presenting them all.}
The algebraic part
for these correlation functions was already calculated in our previous
papers \cite{BGKS07} and \cite{BDGKSW08}. Here we merely cite those
results. For the nearest neighbour correlation functions the following equations
were obtained
\begin{subequations}
\label{corrfncts2}
\begin{align}
     \left\langle\sigma_1^z \sigma_2^z\right\rangle_{T, h} =
        & \cth (\eta) \omega + \frac{\omega'_x}{\eta} \epc\\
     \left\langle\sigma_1^x \sigma_2^x\right\rangle_{T, h} =
        & - \frac{\omega}{2 \sh(\eta)}
          - \frac{\ch(\eta) \omega'_x}{2 \eta} \epp
\end{align}
\end{subequations}
The expansions for next-nearest neighbours read
\begin{subequations}
\begin{align}
\label{eq:z1z3}
\left\langle\sigma^z_1 \s^z_3\right\rangle_{T, h} =&
        2 \cth(2\eta) \omega
        + \frac{\omega^{\prime}_x}{\eta}
	+ \frac{\tgh(\eta) (\omega_{xx} - 2 \omega_{xy})}{4}
        - \frac{\sh^2(\eta) \omega^{\prime}_{xxy}}{4 \eta} \epc \\
      \left\langle\s^x_1 \s^x_3 \right\rangle_{T,h} =&
        - \frac{1}{\sh(2 \eta)} \omega
        - \frac{\ch(2\eta)}{2\eta} \omega^{\prime}_x
%	  \notag \\ & \mspace{180.mu}
	- \frac{\ch(2\eta) \tgh(\eta) (\omega_{xx} - 2 \omega_{xy})}{8}
	+ \frac{\sh^2(\eta) \omega^{\prime}_{xxy}}{8\eta} \epp
\end{align}
\end{subequations}
The length of the formulae grows rapidly with the number of lattice sites.
For $n = 4$ it reads
\begin{multline} \label{szsz4}
     \left\langle\s_1^z \s_4^z\right\rangle_{T, h} =
     \frac{1}{768 q^4 \left(- 1 + q^6 \right) \left(1+q^2\right)
              \eta ^2} \\[1ex]
     \biggl\{ 384 q^4 \left(1+q^2\right)^2
                    \left(5-4 q^2+5 q^4\right) \eta ^2 \omega
		    \mspace{210.mu}\\
     -8 \left(1+q^4 \left(52+64 q^2-234 q^4+64 q^6+52 q^8
               +q^{12}\right)\right) \eta ^2 \omega_{xy} \\
     + 192 q^4 \left(-1+q^2\right)^2
        \left(1+4 q^2+q^4\right) \eta ^2 \omega_{yy} \\
     + \left(-1+q^2\right)^4 \left(1+q^4\right) \left(1+4 q^2+q^4\right)
       \eta^2 \Bigl[ -4 \omega_{xyyy} + 6 \omega_{xxyy} \Bigr] \\[1ex]
     - 768 q^4 \left(-1-q^2+q^6+q^8\right) \eta {\omega'}_y \\
     +16 \left(-1+q^2\right)^3
         \left(1+6 q^2+11 q^4+11 q^6+6 q^8+q^{10}\right)
         \eta  {\omega'}_{xyy}\\
     -2 \left(-1+q^2\right)^5 \left(1+2 q^2+2 q^4+q^6\right)
        \eta  {\omega'}_{xxyyy} \displaybreak[0] \\
     + 8 \left(-1+q^2\right)^3 \left(1+q^2\right)
         \left(1+6 q^2+34 q^4+6 q^6+q^8\right) \eta ^2
	 \Bigl[\omega_y^2 - \omega \omega_{xy} \Bigr] \\
     + \left(-1-4 q^2-22 q^4-12 q^6+12 q^{10}+22 q^{12}
             +4 q^{14}+q^{16}\right) \eta ^2
     \Bigl[ - 6 \omega_{yy}^2 \\
     +12 \omega_{yy} \omega_{xy} + 4 \omega_y \omega_{yyy}
     -12 \omega_y \omega_{xyy} -4  \omega \omega_{xyyy}
     +6  \omega \omega_{xxyy} \Bigr] \displaybreak[0] \\
     + 16 \left(-1+q^2\right)^4 \left(1+q^2\right)^2
          \left(1+q^2+q^4\right) \eta
     \Bigl[ \omega_{yy} {\omega'}_y - \omega_y {\omega'}_{yy} + \omega {\omega'}_{xyy}
     \Bigr] \\
     + \left(-1+q^4\right)^2 \left(1+5 q^2+6 q^4+5 q^6+q^8\right) \eta
       \Bigl[
     4 \omega_{xyyy} {\omega'}_y -6 \omega_{xxyy} {\omega'}_y
     -2 \omega_{yyy} {\omega'}_{yy} \\
     +6 \omega_{xyy} {\omega'}_{yy} +2 \omega_{yy} {\omega'}_{yyy}
     -4 \omega_{xy} {\omega'}_{yyy} -6 \omega_{yy} {\omega'}_{xyy}
     +4 \omega_y {\omega'}_{xyyy} -2 \omega {\omega'}_{xxyyy}\Bigr]\\
     + 3 \left(-1+q^4\right)^3 \left(1+q^2+q^4\right)
     \Bigl[ {\omega'}_{yyy} {\omega'}_{xyy} - {\omega'}_{yy} {\omega'}_{xyyy}
     + {\omega'}_y {\omega'}_{xxyyy} \Bigr] \biggr\}
\end{multline}
and in the transversal case,
\begin{multline} \label{sxsx4}
     \left\langle\s_1^x \s_4^x\right\rangle_{T, h} =
     \frac{1}{3072 q^5
     \left(-1+q^6\right) \eta ^2} \\
     \biggr\{
     -768 q^6 \left(1+10 q^2+q^4\right) \eta^2 \omega \mspace{270.mu} \\
     +16 q^2 \left(-1+q^2\right)^2
        \left(31+56 q^2-30 q^4+56 q^6+31 q^8\right) \eta ^2 \omega_{xy}\\
     -96 q^2 \left(-1+q^2\right)^2
      \left(3+5 q^2-4 q^4+5 q^6+3 q^8\right) \eta ^2 \omega_{yy} \\
     + q^2 \left(-1+q^2\right)^4 \left(1+4 q^2+q^4\right)
     \eta^2 \Bigl[ 8 \omega_{xyyy} -12 \omega_{xxyy} \Bigr]\\
     + 192 q^2 \left(-3-q^2-q^4+q^8+q^{10}+3 q^{12}\right) \eta {\omega'}_y \\
     +8 \left(-1+q^2\right)^3
        \left(1-12 q^2-25 q^4-25 q^6-12 q^8+q^{10}\right)
	\eta  {\omega'}_{xyy}\\
     +2 \left(-1+q^2\right)^5 \left(1+2 q^2+2 q^4+q^6\right)
        \eta {\omega'}_{xxyyy} \displaybreak[0] \\
     +16 q^2 \left(-1+q^2\right)^3 \left(17+7 q^2+7 q^4+17 q^6\right)
        \eta^2 \Bigl[ \omega \omega_{xy} - \omega_y^2 \Bigr] \\
     + q^2 \left(-5-4 q^2-13 q^4+13 q^8+4 q^{10}+5 q^{12}\right) \eta ^2
     \Bigl[ 12 \omega_{yy}^2 -24 \omega_{yy} \omega_{xy} \\
     -8 \omega_y \omega_{yyy} +24 \omega_y \omega_{xyy} +8 \omega \omega_{xyyy}
     -12 \omega \omega_{xxyy} \Bigr]\\
     + 8 \left(-1+q^2\right)^4 \left(1-9 q^2-8 q^4-9 q^6+q^8\right) \eta
     \Bigr[ \omega_{yy} {\omega'}_y - \omega_y {\omega'}_{yy}
           + \omega {\omega'}_{xyy} \Bigr] \displaybreak[0] \\
     + \left(-1+q^4\right)^2 \left(1+5 q^2+6 q^4+5 q^6+q^8\right) \eta
     \Bigl[ -4 \omega_{xyyy} {\omega'}_y +6 \omega_{xxyy} {\omega'}_y\\
     +2 \omega_{yyy} {\omega'}_{yy} -6 \omega_{xyy} {\omega'}_{yy}
     -2 \omega_{yy} {\omega'}_{yyy} +4 \omega_{xy} {\omega'}_{yyy}
     +6 \omega_{yy} {\omega'}_{xyy} -4 \omega_y {\omega'}_{xyyy}
     +2 \omega {\omega'}_{xxyyy} \Bigr] \displaybreak[0] \\
     +3 \left(-1+q^4\right)^3 \left(1+q^2+q^4\right)
     \Bigl[- {\omega'}_{yyy} {\omega'}_{xyy} + {\omega'}_{yy} {\omega'}_{xyyy}
     - {\omega'}_y {\omega'}_{xxyyy} \Bigr] \biggl\} \epc
\end{multline}
with $q=e^{\eta}$.

\begin{figure}[p]
\includegraphics{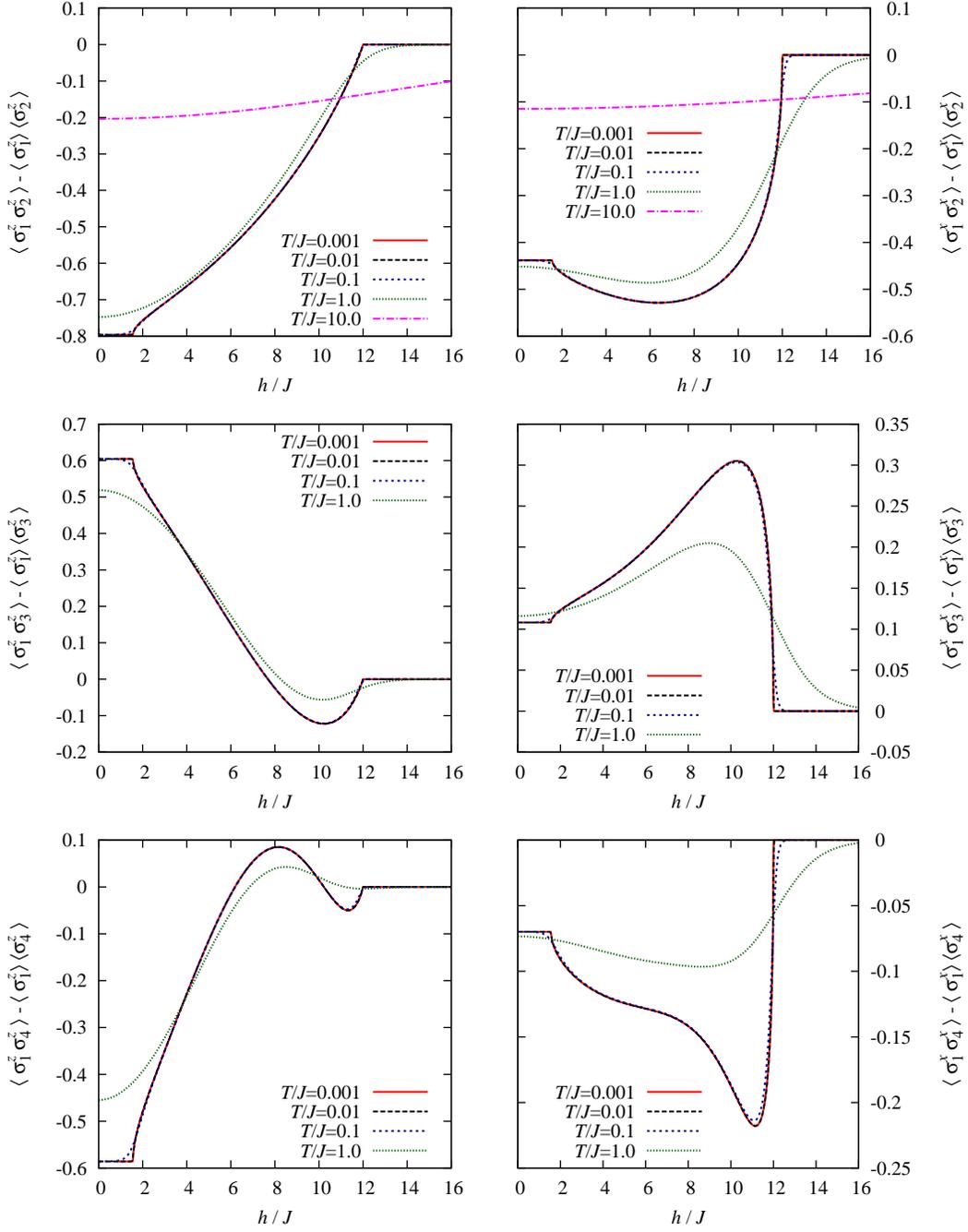}
\caption{(Color online) $\langle\sigma^z_1 \sigma^z_n\rangle
-\langle\sigma^z_1\rangle\langle\sigma^z_n\rangle$ and
$\langle\sigma^x_1 \sigma^x_n\rangle-\langle\sigma^x_1\rangle
\langle\sigma^x_n\rangle$ for different values of
temperature $T/J$ and fixed anisotropy $\Delta=2$. The rows are for
$n=2,3,4$.}
\label{fig:corr_von_h_normalized}
\end{figure}

\begin{figure}[p]
\includegraphics{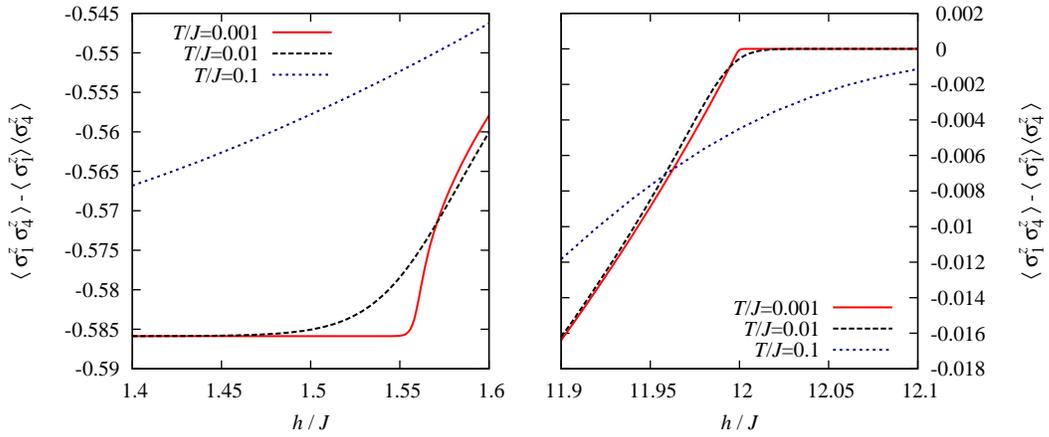}
\caption{(Color online) $\left<\sigma^z_1 \sigma^z_4\right>
-\left<\sigma^z_1\right>\left<\sigma^z_4\right>$ for
different low temperatures $T/J$ and fixed anisotropy $\Delta=2$.
The values of the magnetic field are in the vicinity of the critical
fields.}
\label{fig:corr_von_h_normalized_fein}
\end{figure}

Using the representations via linear and non-linear integral equations
for the functions $\omega$ and $\omega'$ of the previous section we can
determine high-precision numerical values for the various two-point
correlators. Figures \ref{fig:corr_von_h_normalized}-%
\ref{fig:corr_von_h_normalized_fein} show selected examples of connected
two-point functions. The longitudinal correlation functions
$\left\langle\sigma_1^z \sigma_n^z\right\rangle_{T, h}
-\left\langle\sigma_1^z\right\rangle_{T, h}
\left\langle\sigma_n^z\right\rangle_{T, h}$ ($n=2,3,4$) are shown in
the left panels, while the right panels show the transversal correlation
functions $\left\langle\sigma_1^x \sigma_n^x\right\rangle_{T, h}
-\left\langle\sigma_1^x\right\rangle_{T, h}
\left\langle\sigma_n^x\right\rangle_{T, h}$. Note that the
contributions from the one-point functions are somewhat trivial. Due to
the translational invariance of the Hamiltonian the longitudinal
one-point functions do not depend on the site index,
$\left\langle\sigma_1^z\right\rangle_{T, h}=
\left\langle\sigma_n^z\right\rangle_{T, h}$, and are given by
\eqref{eq:phi} for all $n$. The conservation of the $z$-component of
the total spin, on the other hand, implies that
$\left\langle\sigma_n^x\right\rangle_{T, h} =0$. The above notation for
the connected two-point functions is merely used for systematic reasons.

\begin{figure}[p]
\begin{center}
\includegraphics{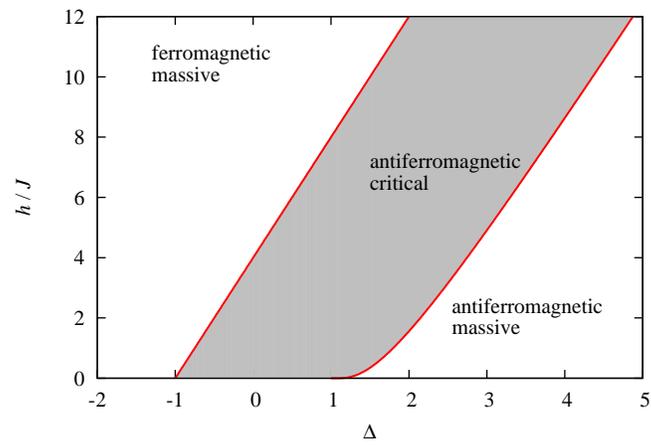}
\end{center}
\caption{(Color online) Ground state phase diagram of the XXZ chain.}
\label{fig:phasendiagramm_xxz}
\end{figure}

In figure \ref{fig:corr_von_h_normalized} we show the dependence of the
correlation functions on the magnetic field for $\Delta=2$ and for several
values of the temperature. At low temperatures one can clearly see the
two critical fields of the ground state phase diagram, figure
\ref{fig:phasendiagramm_xxz}. The saturation field is at $h=12$ and
the critical field at which the excitation gap opens can be calculated
\cite{DeGa66,YaYa66d} to be located at $h\approx 1.55921$. This
can be seen even better in figure \ref{fig:corr_von_h_normalized_fein},
where the $\left\langle\sigma_1^z \sigma_4^z\right\rangle_{T, h}-
\left\langle\sigma_1^z\right\rangle_{T, h}
\left\langle\sigma_4^z\right\rangle_{T, h}$ correlation is shown for
different low temperatures and for magnetic fields in the vicinity of the
two critical fields. Here the curves for $T/J=0.01$ and $T/J=0.001$
are clearly distinguishable, which is not the case on the larger
scale of figure \ref{fig:corr_von_h_normalized}. For fields above the
saturation field $h=12$ the correlators have a fixed value as saturation
sets in. For $h$ below the lower critical field we observe a constant
behaviour as well, which in this case is due to the excitation gap. In
between, however, for values of the magnetic field for which the system
is critical, an interesting non-monotonic behaviour can be seen. It is
most pronounced for the correlators $\left\langle\sigma_1^z
\sigma_4^z\right \rangle_{T, h} -\left\langle\sigma_1^z\right\rangle_{T, h}
\left\langle\sigma_4^z\right\rangle_{T, h}$ for which two local extrema
exist if the temperature is sufficiently small. If the temperature is
too high, thermal fluctuations dominate and all correlations die out.
In figure \ref{fig:corr_von_h_normalized} this is illustrated with the
curve for $T/J=10$ in the $n=2$ case. For $n = 3, 4$ the effect is
similar.

\begin{figure}[p]
\includegraphics{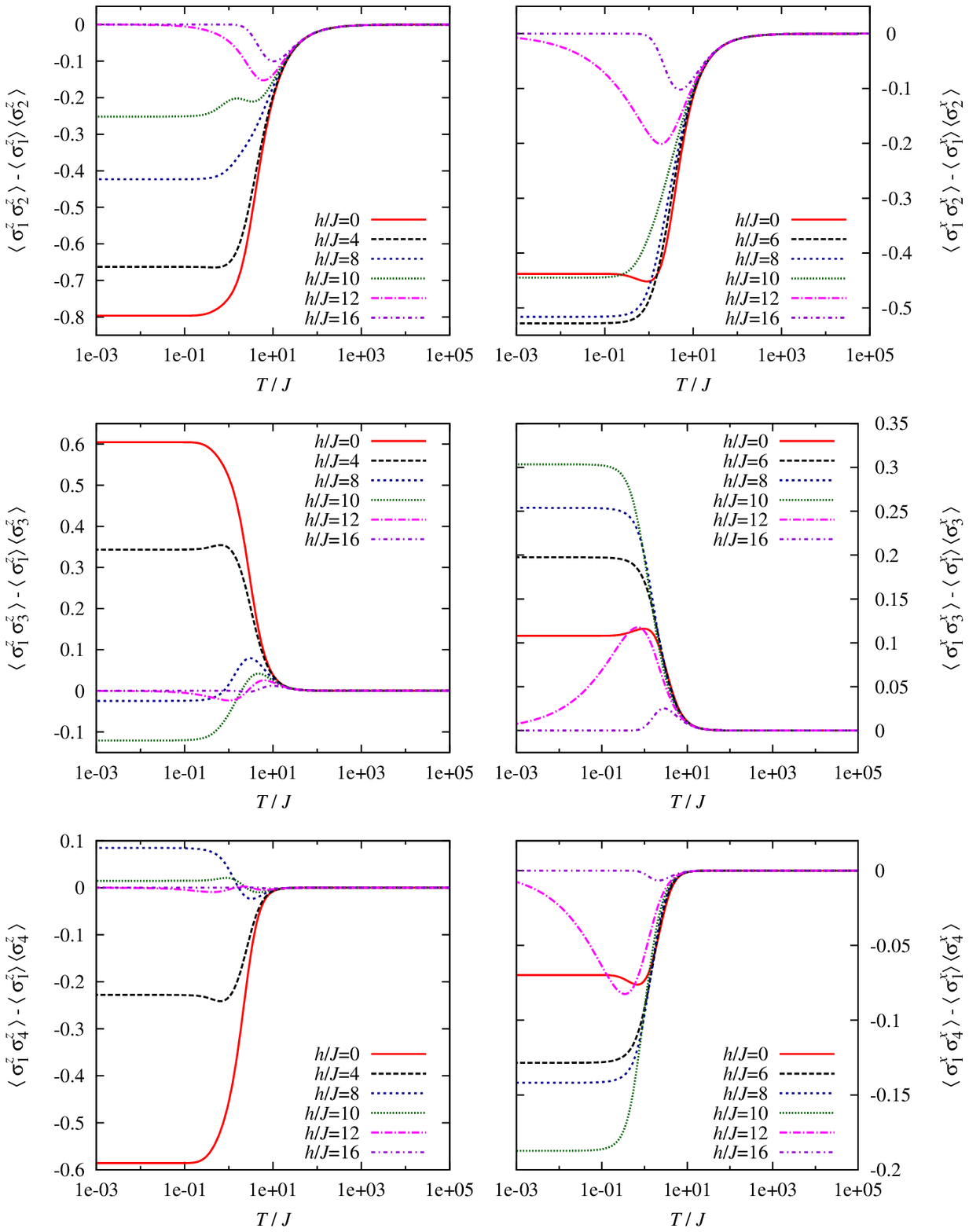}
\caption{(Color online) $\langle\sigma^z_1 \sigma^z_n\rangle
-\langle\sigma^z_1\rangle\langle\sigma^z_n\rangle$ and
$\langle\sigma^x_1 \sigma^x_n\rangle-\langle\sigma^x_1\rangle
\langle\sigma^x_n\rangle$ for different values of the
magnetic field $h/J$ and fixed anisotropy $\Delta=2$. The rows are
for $n=2,3,4$.}
\label{fig:corr_von_T_normalized}
\end{figure}

In figure \ref{fig:corr_von_T_normalized} we show the correlation functions
for several values of the magnetic field as functions of the temperature.
Here we observe a non-monotonic behaviour as well for intermediate magnetic
fields. Regarding the low temperature behaviour we notice that the
connected correlation functions depend in a non-trivial way on the distance
and on the magnetic field. There exists a field for which the connected
correlation functions have maximal modulus. For nearest neighbours
the figures for the transversal correlation functions on the right side,
show the largest values for $h/J=6$, but for next nearest and next-to-next
nearest neighbours the correlations for $h/J=10$ are more pronounced at low
temperature. For the field strength $h/J=16$, which is above
the saturation field, the connected correlation functions deviate the
lesser from zero at intermediate temperatures the larger the distance is.

\begin{figure}[p]
\includegraphics{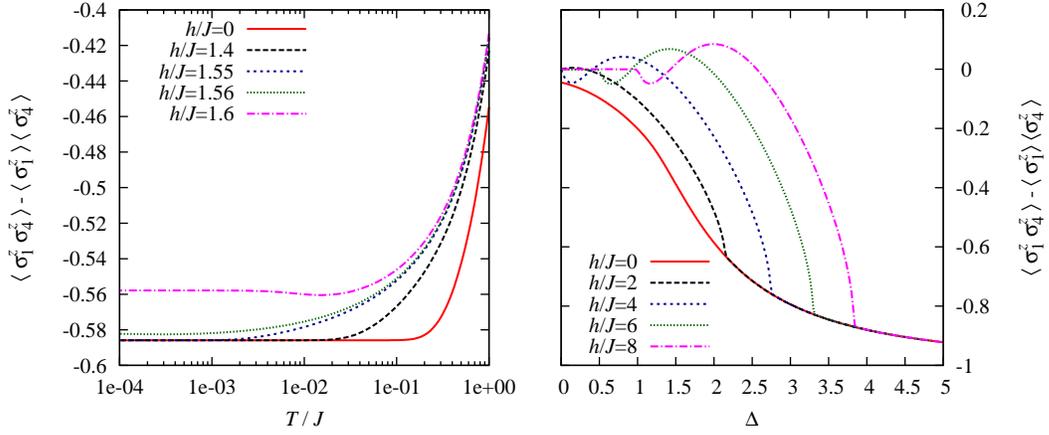}
\caption{(Color online) The left panel shows $\langle\sigma^z_1
\sigma^z_4\rangle-\langle\sigma^z_1\rangle\langle\sigma^z_4\rangle$
for different magnetic fields $h/J$ and fixed anisotropy $\Delta=2$
at low temperatures. The right panel shows $\langle\sigma^z_1
\sigma^z_4\rangle-\langle\sigma^z_1\rangle\langle\sigma^z_4\rangle$
for different magnetic fields $h/J$ as a function of the
anisotropy $\Delta$ at $T/J=0.01$.}
\label{fig:corr_von_T_normalized_fein}
\end{figure}

The excitation gap can also clearly be seen in the
left panel in figure \ref{fig:corr_von_T_normalized_fein}, where we
show $\left\langle\sigma_1^z \sigma_4^z\right\rangle_{T, h}
-\left\langle\sigma_1^z\right\rangle_{T, h}\left\langle\sigma_4^z
\right\rangle_{T, h}$ for low temperatures and several magnetic fields
in the vicinity of the lower critical field. With increasing magnetic
field the correlations start deviating from the $h=0$ line at lower
and lower temperatures, until finally the zero temperature limit
is different from the zero field case. Close to the critical field
the system is particularly sensitive to small changes in the field
strength, which can be well observed by comparing e.g.\ the correlation
functions for $h=1.56$, slightly above the critical field, and
$h=1.55$, slightly below the critical field.

The right panel again shows $\left\langle\sigma_1^z \sigma_4^z
\right\rangle_{T, h}-\left\langle\sigma_1^z\right\rangle_{T, h}
\left\langle\sigma_4^z\right\rangle_{T, h}$, but this time
as a function of $\Delta$ for fixed small temperature $T/J=0.01$ and several
magnetic fields. For large values of $\Delta$ the connected correlation
function is independent of the magnetic field due to the excitation gap.
The effect of the phase transition from the massive fully polarized
state to the critical antiferromagnet is also visible in this figure.
For $h/J=8$, for instance, we observe an abrupt change at $\Delta=1$,
which, at this field strength, is the critical anisotropy at which
saturation sets in. Note that for producing the data for $0 < \Delta < 1$
in the figure we used the formulation and computer implementation
of our previous work \cite{BDGKSW08}. The curves smoothly match at
$\Delta = 1$.

\clearpage

\section{Ising limit}
The subject of this work is the study of correlation functions of
the XXZ chain for $\Delta > 1$. This includes the two extreme cases
$\Delta = 1$ and $\Delta = \infty$. The physically interesting
isotropic point $\Delta = 1$ is difficult to access analytically. In
fact, the regularization of the density matrix by a disorder field
mentioned in the introduction fails exactly at this point, and it
seems that the picture of factorization described above needs a
slight modification in the presence of a magnetic field, when
additional independent functions, called `moments' in \cite{BGKS06},
appear. On the other hand, the numerics discussed in the previous section
remains remarkably stable, if one approaches the isotropic point from
above, and smoothly matches the numerics for approaching it from below
\cite{BDGKSW08}.

For $\Delta = \infty$ or, more precisely, in the limit
$\Delta\rightarrow\infty$ for finite $J_I=J\Delta$ the XXZ Hamiltonian
(\ref{ham}) turns into the Hamiltonian of the Ising chain
\begin{equation}
     \mcH_I=J_I\sum_{j=-N+1}^N\left(\sigma_{j-1}^z\sigma_{j}^z-1\right)
        -\frac{h}{2}\sum_{j=-N+1}^N \sigma_j^z \epp
\end{equation}
The Ising chain may be viewed as a classical one-dimensional model
of statistical mechanics with $2 \times 2$-transfer matrix (see
(\ref{tising}) below). For this reason its correlation functions can
be calculated explicitly \cite{BaBook},
\begin{subequations}
\label{eq:ising_corr_and_mag}
\begin{align}
     \left\langle \sigma_j^z \right\rangle =
        &\frac{\sh\left(\frac{h}{2T}\right)}
              {\sqrt{\sh^2\left(\frac{h}{2T}\right)+e^{4J_I/T}}} \epc\\
               \label{eq:ising_corr}
     \left\langle \sigma_j^z \sigma_k^z \right\rangle =
        & \frac{\sh^2\left(\frac{h}{2T}\right)}
               {\sh^2\left(\frac{h}{2T}\right)+e^{4J_I/T}}
         +\frac{e^{4J_I/T}}{\sh^2\left(\frac{h}{2T}\right)
         +e^{4J_I/T}}\left(\frac{\ch\left(\frac{h}{2T}\right)
         -\sqrt{\sh^2\left(\frac{h}{2T}\right)
                +e^{4J_I/T}}}{\ch\left(\frac{h}{2T}\right)
         +\sqrt{\sh^2\left(\frac{h}{2T}\right)+e^{4J_I/T}}}\right)^{k-j}
          \epp
\end{align}
\end{subequations}

An interesting and non-trivial test of the formulae of the previous section
is to reproduce these results analytically and numerically in the
limit $\Delta \rightarrow \infty$. A similar exercise starting from
the multiple integral formulae of \cite{GKS04a} was carried out in
\cite{GoSe05}. In our case at hand we start with the auxiliary functions
in the $\mfb \mfbq$-formulation which turn into
\begin{subequations}
\label{eq:bbq_ising}
\begin{align}
     \mfb(x) = &e^{-(h/2+4J_I)/T}\left(-\sh\left(\frac{h}{2T}\right)
                +\sqrt{\sh^2\left(\frac{h}{2T}\right)+e^{4J_I/T}}\right)\\
     \mfbq(x) = &\frac{e^{h/2T}}{-\sh\left(\frac{h}{2T}\right)
                +\sqrt{\sh^2\left(\frac{h}{2T}\right)+e^{4J_I/T}}}
\end{align}
\end{subequations}
in the Ising limit. Note that, in this special case, they do not
depend on the spectral parameter. This is also true for the functions
\begin{equation}
     g^\pm_\mu(x) =
        -1\pm\frac{\sh\left(\frac{h}{2T}\right)}
                  {\sqrt{\sh^2\left(\frac{h}{2T}\right)+e^{4J_I/T}}}
\end{equation}
which moreover do not depend on $\mu$ anymore. The same is therefore
true for
\begin{equation}
     \varphi(\mu) = -\frac{\sh\left(\frac{h}{2T}\right)}
                          {\sqrt{\sh^2\left(\frac{h}{2T}\right)
			   +e^{4J_I/T}}}
\end{equation}
and
\begin{equation}
     \omega(\mu_1,\mu_2) =
        -\cth\left(\frac{2J_I}{T}\right)
        +\frac{e^{2J_I/T}}{\sh\left(\frac{2J_I}{T}\right)}
         \frac{\ch\left(\frac{h}{2T}\right)}
              {\sqrt{\sh^2\left(\frac{h}{2T}\right)+e^{4J_I/T}}} \epp
\end{equation}
Similarly, for ${g'}^\pm_\mu$, for which a rescaling is necessary to
obtain a sensible limit, one obtains
\begin{equation}
     \frac{{g'}^\pm_\mu(x)}{\eta} =
        \mp\frac{1}{2}\left(1
        +\frac{\ch\left(\frac{h}{2T}\right)}
              {\sqrt{\sh^2\left(\frac{h}{2T}\right)+e^{4J_I/T}}}\right)
	      \epp
\end{equation}
Note that this rescaling is unproblematic for $\eta\rightarrow\infty$
due to \eqref{eq:gprime}. Thus, we obtain
\begin{equation}
     \frac{{\omega'}(\mu_1,\mu_2)}{\eta}=0 \epp
\end{equation}

The above leads to the correct results for $\left\langle \sigma_1^z
\right\rangle$ and $\left\langle \sigma^z_1 \sigma_2^z \right\rangle$.
For the next-to-nearest neighbour and all higher correlation functions,
however, some of the coefficients in the factorized form of the
correlation functions may diverge in the Ising limit, e.g.\ the term
$\sh^2 (\eta)/\eta$ on the right hand side of \eqref{eq:z1z3}. All the
terms where such type of divergence occurs have to be handled separately.

Sticking with the example of equation \eqref{eq:z1z3} we define
\begin{subequations}
\begin{align}
     f^\pm(x) = &\lim_{\eta\rightarrow\infty}\sh(\eta)\partial_\mu^2
                 \left.g^\pm_{\mut}(x)\right|_{\mu=0}\\[1ex]
     h^\pm(x) = &\lim_{\eta\rightarrow\infty}\sh(\eta)\partial_\mu^2
                 \frac{{g'}^\pm_{\mut}(x)}{\eta}\biggr|_{\mu=0}
\end{align}
\end{subequations}
implying
\begin{subequations}
\begin{align}
     f^+(x) = &-\frac{4}{1+\mfbq}e^{i2x}-4e^{-i2x}\\
     f^-(x) = &-e^{i2x}-\frac{4}{1+\mfb}e^{-i2x}\\
     h^+(x) = &-\frac{4}{1+\mfbq}e^{i2x}\\
     h^-(x) = &\frac{4}{1+\mfb}e^{-i2x}
\end{align}
\end{subequations}
with $\mfb$, $\mfbq$ according to \eqref{eq:bbq_ising}. Using the latter
expressions we conclude that
\begin{equation}
     \lim_{\eta\rightarrow\infty}\frac{\sh^2(\eta)}{\eta}\omega'_{xxy}=
        4-\frac{16}{(1+\mfb)(1+\mfbq)}
\end{equation}
which indeed reproduces \eqref{eq:ising_corr} when inserted into the
equation \eqref{eq:z1z3} for $\left\langle \sigma^z_1 \sigma_3^z
\right\rangle$.

\begin{figure}[ht]
\includegraphics{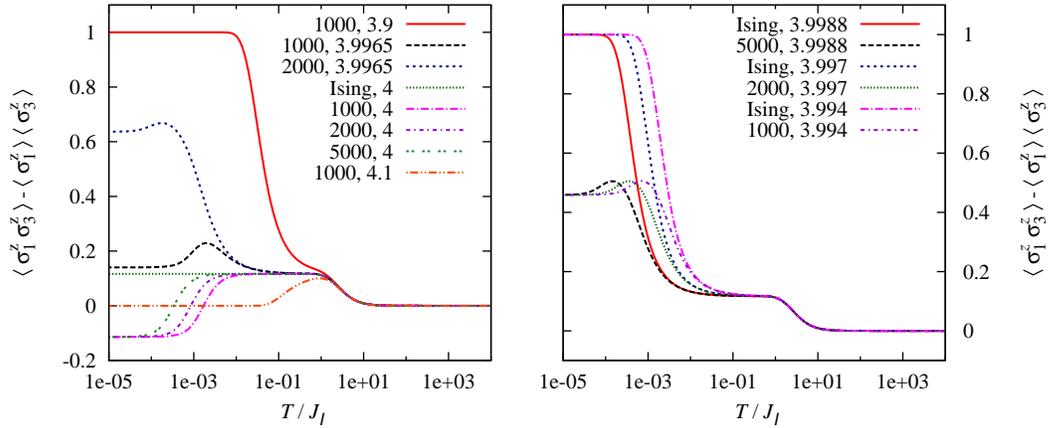}
\caption{(Color online) $\left<\sigma^z_1 \sigma^z_3\right>
-\left<\sigma^z_1\right>\left<\sigma^z_3\right>$ for large values of the anisotropy, fixed $J_I$ and different values of $h/J_I$. The labels in both panels are the tupels $\Delta, h/J_I$ where `Ising' denotes the analytic Ising curves.}
\label{fig:ising}
\end{figure}

After having shown that the Ising limit is included in our representation
of the correlation functions by means of solutions of NLIE, it appears
natural to compare the analytic limit with numerical results for large
$\Delta$. In figure \ref{fig:ising} we show the temperature dependence
of the connected two-point function $\left<\sigma^z_1 \sigma^z_3\right>
- \left<\sigma^z_1\right> \left<\sigma^z_3\right>$ for the Heisenberg
chain near the Ising limit, for different values of the anisotropy
parameter and the magnetic field. In general, these curves do not depend on
$\Delta$ for large anisotropies and match the curves for the Ising limit
given by \eqref{eq:ising_corr_and_mag}. E.g., the curves in the left panel
for $\Delta = 1000$ and $h/J_I=3.9$ or $h/J_I=4.1$, respectively,
match the corresponding Ising curves to the precision of the line width
in the figure. In the vicinity of $h = 4J_I$, however, we observe large
deviations from the low-temperature Ising curves.

\begin{figure}[p]
\begin{center}
\includegraphics{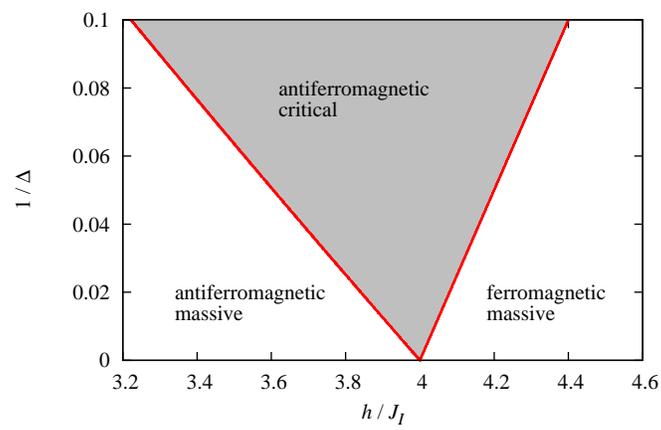}
\end{center}
\caption{(Color online) Ground state phase diagram of the XXZ chain
in the $h$-$1/\Delta$ plane for fixed $J_I=J\Delta$ and large $\Delta$.}
\label{fig:phasendiagramm_ising}
\end{figure}

This behaviour is induced by a critical point at $1/\Delta = 0$,
$h = 4J_I$ in the zero temperature phase diagram of the XXZ chain
(see figure \ref{fig:phasendiagramm_ising}) which separates regions
of different asymptotics for the two-point correlation functions
of the Ising chain. The latter fact is most easily understood by
inspection of the $2 \times 2$ transfer matrix of the Ising chain
\cite{BaBook},
\begin{equation} \label{tising}
     t = \left(\begin{array}{cc}e^{h/2T} & e^{2J_I/T}\\
        e^{2J_I/T}&e^{-h/2T}\end{array}\right) \epp
\end{equation}
Its eigenvalues $\lambda_\pm$ ($\lambda_+ > \lambda_-$) and eigenvectors
$\opv_\pm$ determine the partition function of the Ising chain as well
as its two-point functions. In fact, when we wrote \eqref{eq:ising_corr},
we used the formula
\cite{BaBook}
\begin{equation} \label{isinggenform}
\left\langle \sigma_1^z \sigma_{n+1}^z \right\rangle = 
\left\langle \opv_+,\sigma^z \opv_+\right\rangle^2 +
\left\langle \opv_+,\sigma^z \opv_-\right\rangle^2 \left(\frac{\lambda_-}{\lambda_+}\right)^n
\end{equation}
and the explicit expression for the eigenvalues and eigenvectors of $t$,
which the reader may easily calculate from (\ref{tising}). In order
to understand the different zero temperature behaviour of the Ising
correlation functions it suffices to consider the low temperature
asymptotics of the transfer matrix. Using (\ref{tising}),
(\ref{isinggenform}) one easily distinguishes three different asymptotic
regimes.
\begin{subequations}
\begin{align}
     \bullet \quad h>4J_I: & &
     t \sim e^{h/2T} &
        \left(\begin{array}{cc} 1 & 0\\ 0 & 0 \end{array}\right) \epc &
        \quad \left\langle \sigma^z_1 \sigma_{n+1}^z \right\rangle \sim 1
        \epc \\[1ex]
     \bullet \quad h<4J_I: & &
     t \sim e^{2J_I/T} &
        \left(\begin{array}{cc} 0 & 1\\ 1 & 0\end{array}\right) \epc &
        \quad \left\langle \sigma^z_1 \sigma_{n+1}^z \right\rangle
        \sim (-1)^n \epc \\[1ex]
     \bullet \quad h=4J_I: & &
     t \sim e^{2J_I/T} &
        \left(\begin{array}{cc} 1 & 1\\ 1 & 0\end{array}\right) \epc &
        \quad \left\langle \sigma^z_1 \sigma_{n+1}^z \right\rangle
        \sim \frac{1}{5}
        +\frac{4}{5}\left(\frac{1-\sqrt{5}}{1+\sqrt{5}}\right)^n \epp
\end{align}
\end{subequations}

Thus, in the Ising limit the absolute value of the ground state correlation
functions is generally independent of the spatial separation, except at
the critical field $h_c = 4 J_I$, where we see exponential decay due to
the residual entropy resulting from an exponential degeneracy of the ground
state. For finite $\Delta$ this degeneracy is lifted by residual quantum
mechanical interactions causing algebraically decaying correlations.
The critical point of the Ising chain corresponding to a first order
phase transition appears as a critical point in the ground state phase
diagram of the XXZ chain if we draw it in the $h$-$1/\Delta$ plane for
fixed $J_I$, see figure \ref{fig:phasendiagramm_ising}.

Contrary to the Ising case, there are always two critical fields if $\Delta$
is finite. They correspond to two second order phase transition lines at the
boundary of the critical phase in the $h$-$1/\Delta$ diagram. This means
that, away from the Ising limit, even the ground state correlation
functions depend continuously on the magnetic field. Hence, for sufficiently
low temperatures, there must be deviations from the Ising curves for
values of $h$ and $\Delta$ belonging to the critical phase. In general
the zero temperature limit should depend on the anisotropy and the
magnetic field. This can be seen in the left panel in figure \ref{fig:ising}
for $h/J_I=3.9965$, $\Delta=1000$ and $\Delta=2000$. However, as the width of
the critical phase gets smaller with increasing $\Delta$, one can find for
each magnetic field, except for the critical field, a sufficiently large
$\Delta$ such that for larger anisotropies the correlation functions
behave as in the Ising model. 

For $h/J_I=4$, there is always a deviation in the zero temperature
behaviour between the Ising chain and the XXZ chain for finite $\Delta$.
See figure \ref{fig:ising}, where the connected correlation functions are
shown for $\Delta=1000$, 2000, 5000 and in the Ising limit. Interestingly
in this case the zero temperature limit does not depend on the (finite)
value of $\Delta$. This is a general behaviour as we observe that the
correlation functions are asymptotically constant on straight lines ending
in the Ising critical point. We can see this exemplarily in the right panel
of figure \ref{fig:ising}, where the connected correlation functions show
the same zero temperature asymptotics for three different pairs of values
of the anisotropy and the magnetic field.

Comparing these curves with those for the Ising model with the same
magnetic fields, we identify four different temperature regimes. For
very high temperatures the curves are independent of $\Delta$ and $h$. Then,
for intermediate temperatures, they are independent of the (large) anisotropy
and only depend on the magnetic field. This is where the curves for
finite $\Delta$ and the curves of the Ising model match. Next comes an
regime where the correlation functions depend on the anisotropy
and the magnetic field, and finally, for very low temperatures, only the
product $(h - h_c) \Delta$ determines the correlation functions. In fact,
this is a rare example, where the full low temperature asymptotics of
thermodynamic quantities and short-range correlation functions
in the vicinity of a critical point can be worked out analytically
\cite{GKT09b}.

\clearpage

\section{Paramagnet}
Another limit, which is interesting from a technical point of view, is
the paramagnet $J=0$. At a first glance it might look rather trivial. Yet,
as the coefficients in section \ref{sec:examples} depend on $\eta$, the
functions $\omega$, $\omega'$ and $\varphi$ depend on $\eta$ as well, but
the final result must not. In this sense the paramagnetic limit is even
more intricate than the Ising limit, as, in contrast to the latter, all
the summands in the algebraic part contribute. This way we have an
additional test for the coefficients in \eqref{corrfncts2}-\eqref{sxsx4}.
We note that the physical part takes the form
\begin{subequations}
\begin{align}
     \varphi(\mu) = & -\tgh\left(\frac{h}{2T}\right) \epc \\
     \omega(\mu_1,\mu_2) = &\tilde{K}_\eta(\mut_2-\mut_1)
                            \tgh^2\left(\frac{h}{2T}\right) \epc \\
     \omega'(\mu_1,\mu_2) = &\tilde{L}_\eta(\mut_2-\mut_1)
                             \tgh^2\left(\frac{h}{2T}\right) \epp
\end{align}
\end{subequations}
in the paramagnetic limit.

\section{Conclusion}
We have studied the short-range correlation functions of the
XXZ chain in the massive phase by means of the equations
(\ref{corrfncts2})-(\ref{sxsx4}) representing them in factorized form.
For this purpose we derived in section \ref{sec:physpart} a
representation of the physical part of the correlation functions which
is well suited for the implementation on a computer. We obtained
high-accuracy data for the correlation functions of the spin chain in
the thermodynamic limit. Together with our previous results
\cite{BDGKSW08} we can now access the full parameter plane of the
infinite antiferromagnetic chain (anisotropy $\Delta > - 1$ and magnetic
field $h$ arbitrary) at arbitrary temperatures. The short-range
correlation functions show a surprisingly rich non-monotonous
behaviour at intermediate magnetic fields and temperatures. At low
enough temperatures the critical lines of the ground state phase
diagram can be read off from our data. The numerics is stable 
in the isotropic limit $\Delta \rightarrow 1+$ and in the Ising limit
$\Delta \rightarrow \infty$.

So far our method is still limited in the accessible range of the
correlation functions. In order to extend this range, an efficient
algorithm for the calculation of the algebraic part of the
correlation functions is needed. Since this is a well defined, purely
algebraic problem though, we have little doubt that it will be solved
in the future. For the exact calculation of the large-distance asymptotics
at finite temperatures, on the other hand, we believe that new insight
will be needed (for recent progress on the corresponding ground state
problem see \cite{KKMST09a}).

\section*{Acknowledgment}
The authors are grateful to M. Karbach for helpful discussions.
CT likes to acknowledge support by the research program of the
Graduiertenkolleg 1052 funded by the Deutsche Forschungsgemeinschaft.

%\bibliographystyle{amsplain} 
%\bibliography{hub_christian}

\providecommand{\bysame}{\leavevmode\hbox to3em{\hrulefill}\thinspace}
\providecommand{\MR}{\relax\ifhmode\unskip\space\fi MR }
\providecommand{\MRhref}[2]{%
  \href{http://www.ams.org/mathscinet-getitem?mr=#1}{#2}
}
\providecommand{\href}[2]{#2}

\end{document}